\documentclass[prl,twocolumn,preprintnumbers,nofootinbib,tightenlines,superscriptaddress]{revtex4}

\usepackage{graphicx}
\usepackage{amssymb}
\usepackage{amsmath,mathrsfs,verbatim}
\usepackage{times}
\usepackage{latexsym}
\def\beq{\begin{equation}}
\def\eeq{\end{equation}}
\def\bey{\begin{eqnarray}}
\def\eey{\end{eqnarray}}

\def\lsim{\mathrel{\raise.3ex\hbox{$<$\kern-.75em\lower1ex\hbox{$\sim$}}}}
\def\gsim{\mathrel{\raise.3ex\hbox{$>$\kern-.75em\lower1ex\hbox{$\sim$}}}}

\newcommand{\be}{\begin{equation}}
\newcommand{\ee}{\end{equation}}

\newcounter{sec}

\begin{document}

\title{A New Challenge for Dark Matter Models}

\author{Mohammadtaher Safarzadeh \& Abraham Loeb\\
Center for Astrophysics | Harvard \& Smithsonian, 60 Garden Street, Cambridge, MA}

\date{\today}

\begin{abstract}
\vspace*{.0in}
Cold dark matter (CDM) has faced a number of challenges mainly at small scales, such as the too-big-to-fail problem, and core-cusp density profile of dwarf galaxies. 
Such problems were argued to have a solution either in the baryonic physics sector or in modifying the nature of dark matter to be self-interacting, or self-annihilating, or ultra-light.
Here we present a new challenge for CDM by showing that two of Milky Way's satellites (Horologium I, and Tucana II) are too dense, requiring the formation masses and redshifts of halos in CDM not compatible with being a satellite. 
These too-dense-to-be-satellite systems are dominated by dark matter and exhibit a surface density above mean dark energy cosmic surface density $\sim\Omega_{\Lambda} \rho_c c/H_0\approx 600~\rm M_{\odot}/pc^2$. 
This value corresponds to dark matter pressure of $\approx 10^{-9}{\rm erg/cm^3}$.
This problem, unlike other issues facing CDM, has no solution in the baryonic sector and none of the current alternatives of dark matter can account for it. 
The too-dense-to-be-satellite problem presented in this work provides a new clue for the nature of dark matter, never accounted for before. Moreover, we find that a number of MW's satellite require formation halo masses below the atomic cooling limit which by itself is another challenging observation to account for in CDM. 

\end{abstract}


\maketitle

\stepcounter{sec}
{\bf Introduction.\;}  
Collisionless dark matter has been a popular candidate for dark matter despite problems at small scales such as the missing satellite problem~\cite{Moore1999ApJ,Klypin1999ApJ}, too-big-to-fail (TBTF)~\cite{tbtf}, core-cusp problem~\cite{WP11,Kroupa2010}.
disk-of-satellites~\cite{Kroupa2005,Pawlowski2020}, and larger scales such as galaxy bars~\cite{Roshan2021}, and galaxy clusters~\cite{Meneghetti2020Sci}. We refer the readers to discussion presented in Ref.~\cite{Kroupa2010,Salucci2019AAPR} for other issues with CDM.

Attempts to rectify these issues can be grouped into either modification to the nature of dark matter~\cite{SS2000PhRvL,Viel2005PhRvD,Schive2014,Vogelsberger2012,Rocha2013MNRAS,Tulin2018PhR}, or modification of baryonic physics feedback~\cite{D'Onghia2010ApJ,PG2012,Governato2012,Sawala2017MNRAS,GarrisonKimmel2019MNRAS}. For a recent review see~\cite{BBK2017}.

So far no single solution has been able to solve all the problems at small scales without creating new problems on larger scales. 
For example, the self-interacting dark matter model can alleviate the too-big-to-fail problem, but not the missing satellite one~\cite{Rocha2013MNRAS}.
Similarly ultra-light dark matter model fails to account for the diversity of dwarf galaxy profiles~\cite{SS2020}. Solutions aiming at missing satellite problems can not simultaneously account for the TBTF problem.

In this {\it Letter}, we explore the constraints on the formation mass and redshift of dark matter halos hosting classical and ultra-faint dwarf galaxies that are satellites of the Milky Way (MW) in the standard cold dark matter (CDM) cosmology.
We show that a single data point of half-light radius and the enclosed density within half-light radius can be compared to NFW profile resulting in unjustifiable formation redshift and halo mass for such dwarf galaxies.
This problem arises since some of the dwarf galaxies are simply too dense to account for in the standard cosmology~\cite{Planck2015}.
Since these are small galaxies without significant baryonic content, adiabatic contraction can not account for the inferred high densities. Even though such high densities are attainable in the core-collapse regime of self-interacting dark matter model in the presence of tidal stripping, the satellites in our sample are prone to tidal stripping due to their estimated orbits from Gaia DR2~\cite{Simon2018ApJ,Hammer2020ApJ,Li2021arXiv210403974L}.

Recently, the Very same problem on has been pointed out in Ref. \cite{Meneghetti2020Sci} on
larger scales, that observed cluster substructures are more efficient lenses, due to excess central surface mass densities than predicted by CDM simulations, by more than an order of magnitude. 
Unlike other issues that have nagged the CDM model, these two findings report an observed enhancement in central densities well in excess of predictions by LCDM. 

\begin{figure*}
\includegraphics[scale=.4]{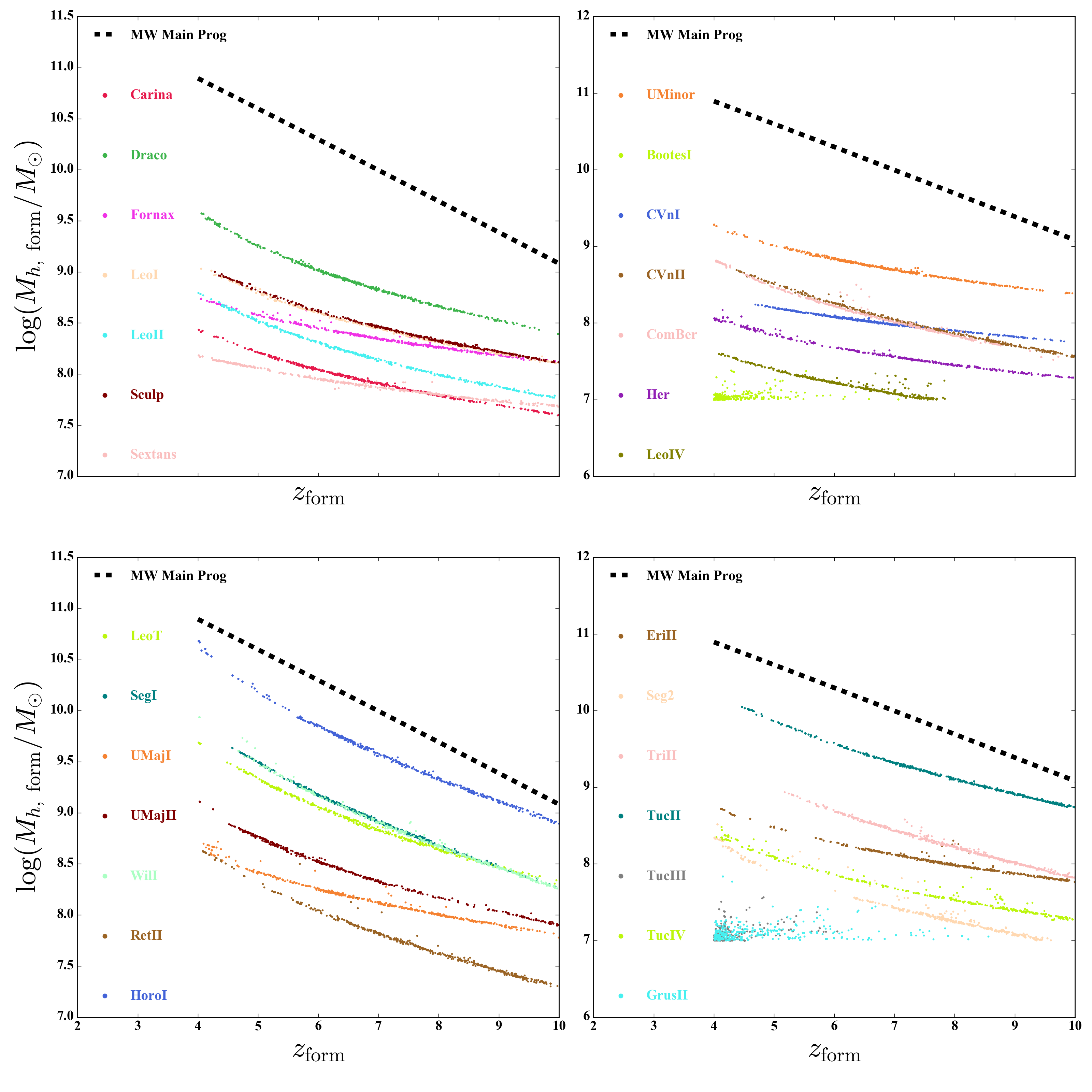} 
\caption{Constraints on formation redshift (horizontal) and correpsonding halo mass (vertical) of classical dwarfs and ultra-faint dwarf galaxies when fitting an NFW profile against a single data point of enclosed mass density within half-light radius. 
The thick black dashed line in all panels shows the MW's mass assembly history, indicating that a satellite of a MW should be far below the black dashed line at all times as otherwise would merge with the main progenitor due to dynamical friction~\cite{Safarzadeh2018MNRAS}.}
\label{fig:fig1}
\end{figure*}

\stepcounter{sec}
{\bf NFW halo model fit\;}  

We compile the density within the half-light radius of both classical dwarf galaxies and ultra-faint dwarf galaxies from Ref.~\cite{Wolf2010} with updated values from the literature as discussed in Ref.~\cite{SL2021}.

We construct Navarro-Frenk-White (NFW)~\cite{NFW1996,NFW1997} profiles for a halo of mass $M_h$, and formation redshift $z_f$ as follows:
\be
\rho(r) = \rho_s (r/r_s)^{-1}(1+r/r_s)^{-2},
\ee
where $\rho_s=\rho_c \delta_c$ with $\rho_c=3H_z^2/8\pi G$ and $\delta_c$ given by:
\be
\delta_c=\frac{200}{3}\frac{c^3}{\ln(1+c)-c/(1+c)}.
\ee
The concentration parameter of a halo depends on its formation mass and redshift and we adopt the derived relationships from N-body simulations from Ref.~\cite{Correa2015} which is in agreement with other studies as compiled in Ref.~\cite{Diemer2019ApJ}.
In these calculations $r_s=r_{200}/c$ where $r_{200}$ is given by:
\be
r_{200}\approx31 (\frac{M_h}{10^{12}M_{\odot}})^{1/3} (\frac{\Omega_m^z}{\Omega_m})^{-1/3} (\frac{1+z}{7})^{-1}{\rm kpc},
\ee
and $\Omega_m^z\approx\Omega_m(1+z)^3/(\Omega_\Lambda+\Omega_m(1+z)^3)$.
For a flat universe, the Hubble parameter evolution is given by:
\be
H_z^2=H_0^2 \big[(1+z)^3 \Omega_m +\Omega_\Lambda)\big],
\ee
and we adopt Planck 2015 cosmology~\cite{Planck2015} with $H_0$=68 ${\rm km/s/Mpc}$, $\Omega_m=0.3$, and $\Omega_\Lambda=0.7$.

We fit the NFW profile with two free parameters of formation redshift and halo mass to the single data point for each galaxy, which is the density within its half-light radius. For all the galaxies we consider a typical 10\% error on the enclosed density. We perform an MCMC calculation by implementing {\rm emcee}~\cite{emcee2013PASP} and the resulting posterior samples are shown in Figure \ref{fig:fig1}. Each galaxy is shown with a different color, and the thick red dashed line in all panels indicates the MW's assembly history~\cite{vdB2002,Fakhouri2010,Correa2015III}, meaning the main progenitor of the MW follows the red dashed line. 
Therefore, a satellite of the MW should be below that line at all redshifts. While a substantial fraction of the satellites falls below the black dashed line, a number of satellites have their formation redshift and halo mass close to the MW's line which would not be consistent with them being a MW satellite as such haloes merge with the main progenitor of the MW due to dynamical frictions and therefore not survive as a satellite~\cite{Safarzadeh2018MNRAS}. We note that similar line of discussion has been discussed previously using different metrics that is adopted in this work~\cite{Geringer2015}.

\begin{figure}
\includegraphics[scale=0.4]{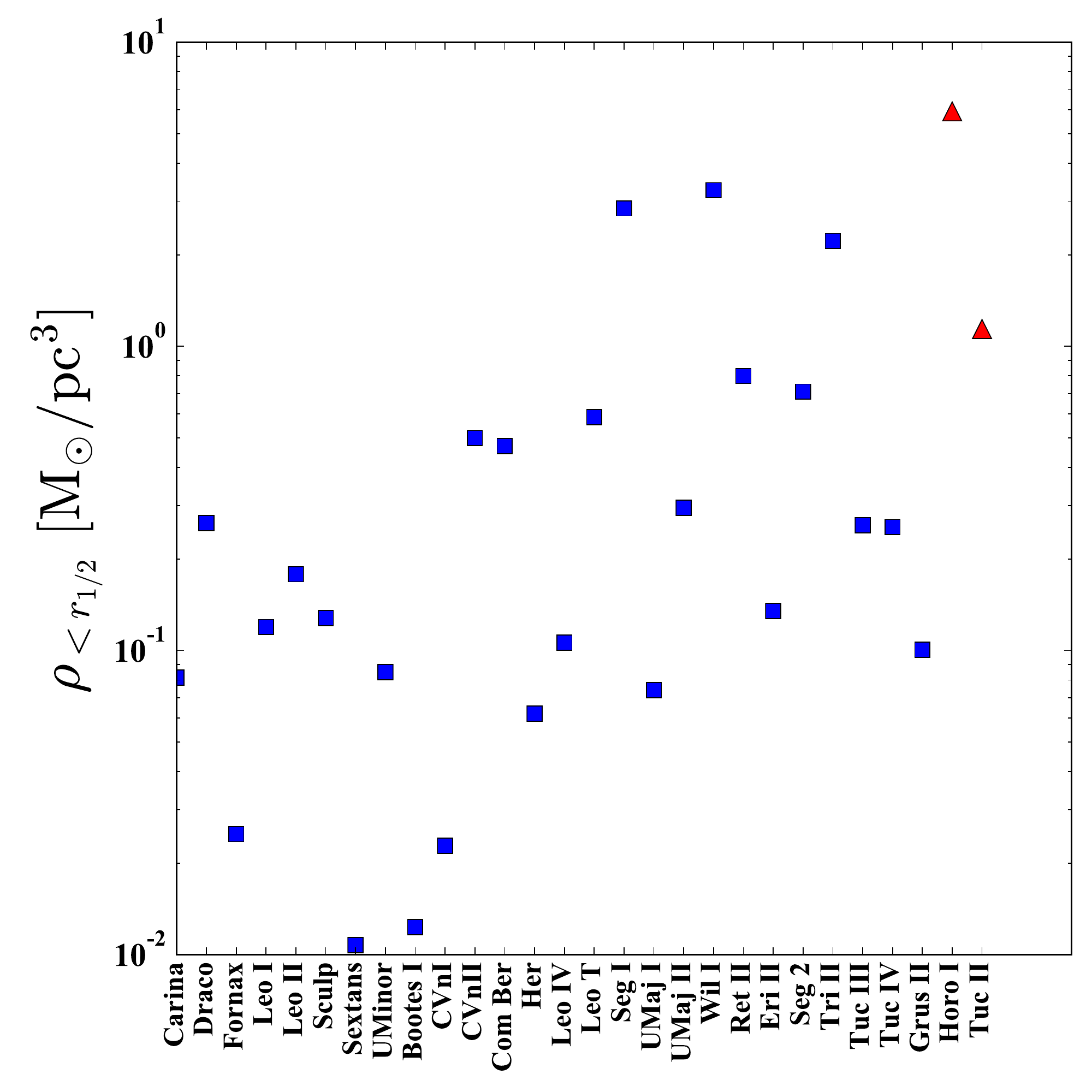} 
\includegraphics[scale=0.4]{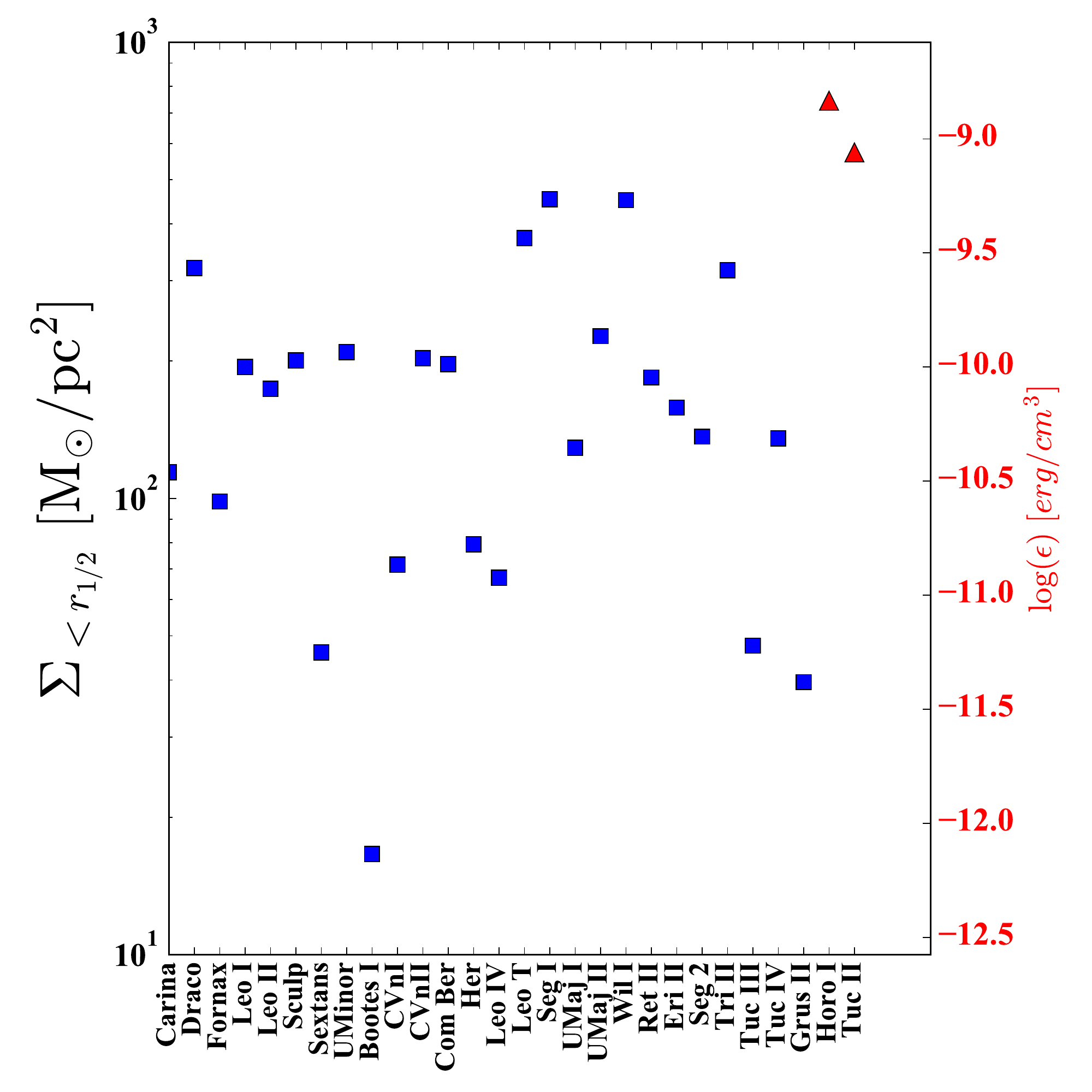} 
\caption{{\emph Upper panel:} Dwarf galaxies with their density within half-light radius. Blue squares are galaxies with their formation redshift and halo mass below the MW's line and the red triangles are galaxies with estimated formation redshift and halo mass at formation close the MW's assembly line. {\emph Bottom panel}: similar to the left panel but showing the surface density within the half-light radius. }
\label{fig:fig2}
\end{figure}


Next, we look for the difference between the halos that lie below and close to the MW assembly history. The left panel of Figure \ref{fig:fig2} shows the dwarf galaxies with their density within the half-light radius. Blue squares are galaxies with their formation redshift and halo mass below the MW's line and the red crosses are galaxies with estimated formation redshift and halo mass at formation close to the MW's assembly line. The right panel of Figure \ref{fig:fig2} is similar to the left panel but showing the surface density within the half-light radius. We can see that 600 $\rm {M_{\odot}/pc^2}$ is the threshold above which satellites will have formation halo mass and redshift inconsistent with being a MW's satellite. 

It is interesting to note that such a threshold for dark matter surface density is similar in magnitude to the cosmic mean surface density of dark energy $\Sigma_{\Lambda}=\Omega_{\Lambda}\rho_c c/H_0$. 
Given that energy density or pressure is related to surface density as $\epsilon=\rho v^2\sim (M/R^3)(GM/R)\sim G \Sigma_{dm}^2$, for a virialized system of mass $M$, radius $R$, and characteristic velocity $v$, we also show the corresponding values for dark matter pressure in Figure \ref{fig:fig2}.
Whether this provides a clue to the nature of dark matter should be further investigated. 

\stepcounter{sec}
{\bf Discussion.\;} 
Unlike the TBTF or core-cusp problems, which require lowering the density in the satellite's core to match the observations, the too-dense-to-be-satellite problem asks for the opposite to take place; namely some of the satellites require a mechanism through which the density within their half-light increase (in the language of NFW profile, requiring a steeper profile) than predicted in CDM cosmology. Can this be accommodated through baryonic physics or modification of dark matter properties?

The only way that baryonic physics can help condense dark matter is through adiabatic compression~\cite{Blumenthal1986ApJ,Gnedin2004ApJ}. 
However, such a mechanism requires baryonic mass densities comparable to dark matter densities which is not the case for the systems we have explored in this work. 
We find no correlation between the ratio of stellar to dark matter mass within the half-light radius and the enhancement of the dark matter surface density. 
The baryonic content of the ultra-faint dwarf galaxies is negligible as these systems are dark matter-dominated systems~\cite{SG2007ApJ,Simon2011ApJSegI,Kirby2017ApJ,Simon2019ARAA}.

Could the enhancement in central surface density be explained by novel dark matter properties? It has been shown that self-interacting dark matter (SIDM) can follow a core-collapse phase if the satellite undergoes tidal disruption and is born with a high concentration parameter~\cite{Sameie2020PhRvL}. However, the estimated orbital parameters for the galaxies analyzed in this work suggest little to no dark matter tidal stripping for them. For example, the galaxy Horologium I has a low eccentricity orbital parameter ($e\approx0.8$), and peri(apo)center distance of $87 (741)$ kpc~\cite{Simon2018ApJ}. The tidal radius at such a distance from the Galactic center is~\cite{Read2006MNRAS}:
\be
r_t\approx r_p \big[\frac{M_s}{M_{MW}(3+e)}\big]^{1/3},
\ee
where $r_p$ is the pericenter distance, and $M_s$ and $M_{MW}$ is the mass of the satellite and MW respectively. Adopting $r_p=87 {\rm kpc}$, and $M_s=10^9~M_{\odot}$ and $M_h=10^{12}~M_{\odot}$ we find $r_t=5.5~{\rm kpc}$.
Comparing this to the core radius of a $10^9~M_{\odot}$ halo formed at redshift $z_f=10$, or  a $10^{10}~M_{\odot}$ halo formed at redshift $z_f=5$, we get $r_{s}=1.02 (3.38)~{\rm kpc}$ respectively.  
We observe that $r_t$ is too large to impact the core of the satellite's host halo. However, alternatives to SIDM might have less issue with regards to lack of tidal stripping~\cite{Essig2019PhRvL}.

Alternatively, the enhanced density of dark matter above a surface density of about 600 $\rm M_{\odot}/pc^2$, or acceleration of $g_{\rm crit}=G \Sigma_{\rm crit}\approx10^{-8} \rm cm/s^2$, or an energy density (pressure) of $10^{-9}\rm erg/cm^3$ may reflect a fundamental clue about the nature of dark matter. We defer the particle physics interpretation to future studies.

Finally we note that our analysis indicates a number of MW's satellites should have been born with halo masses below the atomic cooling limit~\cite{LFbook} which by itself is challenging to account for in LCDM ~\cite{Kirby2013ApJ,Graus2019MNRAS}

{\it Acknowledgments}: We are thankful to James Bullock and Mike Boylan-Kolchin for correcting us regarding an error in our calculations and insightful discussions. We are thankful to Josh Simon, Priyamvada Natarajan, Jerry Ostriker, 
Savvas Koushiappas, Manoj Kaplinghat, and Pavel Kroupa for useful comments. This work was supported in part by the Dean’s Competitive Fund for Promising Scholarship at the Faculty of Arts \& Sciences of Harvard University.

\end{document}